\documentclass
[aps,prb,amsfonts,amssymb,twocolumn,amsmath,preprintnumbers,nofootinbib,floatfix,
showpacs]{revtex4-1}%
\usepackage[dvips]{graphics}
\usepackage{graphicx}
\usepackage{bm}
\usepackage{amsmath}
\usepackage{amsfonts}
\usepackage{amssymb}%
\providecommand{\U}[1]{\protect\rule{.1in}{.1in}}

\begin{document}

\title{Unconventional thermoelectric behaviors and enhancement of figure of merit in
Rashba spintronic systems}
\author{Cong Xiao, Dingping Li, and Zhongshui Ma}
\affiliation{School of Physics, Peking University, Beijing 100871, China \\
Collaborative Innovation Center of Quantum Matter, Beijing, 100871, China}

\begin{abstract}
Thermoelectric transport in strongly spin-orbit coupled two-dimensional Rashba system is
studied using the exact solution of the linearized Boltzmann equation.
Some unusual transport behaviors are revealed.
We show that the electrical conductivity takes a Drude form when the Fermi
energy $E_{F}$ is above the band crossing point, but a non-Drude form which is
a quadratic function of $E_{F}$ when $E_{F}$ lies below the band crossing
point. The Mott relation breaks down when $E_{F}$ lies in the vicinity of
the band crossing point. It is shown that the thermopower and thermoelectric figure
of merit are strongly enhanced when $E_{F}$ downs below the band crossing
point. This enhancement is attributed to not only the one-dimensional-like
density of state but also the unconventional intraband elastic scattering
below the band crossing point. The differences between these results and those obtained
by the relaxation time approximation are discussed in detail.
\end{abstract}

\maketitle
\section{Introduction}

In two-dimensional electron system (2DES) with Rashba spin-orbit coupling
(SOC), when the energy downs below the band crossing point (BCP), a band
valley emerges (Fig. 1) with different topology of Fermi surfaces (FS) from
that when the energy is above the BCP \cite{Cappelluti2007,Tsutsui2012,Lv2013}%
. In the band valley the dispersion curve is not monotonic and the density of
state (DOS) is one-dimensional (1D)-like. In the case of strong SOC, the Fermi
level can lie in the vicinity of or even below the BCP, and the valley
structure can survive the weak disorder and thermal smearing at low
temperatures. Therefore, in this case the nontrivial topology of FS in the
band valley may affect the transport properties significantly.

The strongly spin-obit coupled 2DES, formed at the Te-terminated surface in
layered polar semiconductors BiTeX (X=Cl, Br, I), has been discovered recently
by ARPES measurements in agreement with ab-initio calculations
\cite{Eremeev,Landolt,Crepaldi2012,Sakano2013,Rusinov2013}. In such 2DES the
giant Rashba SOC coefficient is one order of magnitude higher than that in
conventional III-V semiconductor heterostructures \cite{Nitta1997}. In
addition, the BiTeX quantum well \cite{Wu2014b} may be another candidate to
realize strongly spin-orbit coupled 2DES. Very recently the first-principle
band structure calculation has suggested the strain engineering of heavy-metal
film on layered large-gap semiconductor substrate \cite{Ming2015} as a
promising way to form 2DES with strong Rashba SOC, e.g., Au single layer on
strained InSe(0001). In aforementioned 2DES, the large Rashba spin-splitting
provides a chance to study the unconventional transport properties induced by
the band valley and band crossing.

The effects of the band valley and band crossing on spin transport and
superconducting electronics have received much theoretical attention, e.g.,
the non-Dyakonov-Perel spin relaxation behavior \cite{Grimaldi2005}, the
non-Edelstein electric-field induced spin polarization \cite{Dyrdal2013,Ref-2}%
, the enhanced spin-orbit torque efficiency \cite{Tsutsui2012}, the enhanced
superconducting instabilities \cite{Cappelluti2007} and the specular Andreev
reflection in the interface of a superconductor and a 2DES with strong Rashba
SOC \cite{Lv2012}. In addition, spin-related thermoelectric conversion in
systems with strong Rashba SOC is gathering increasing attention, which is not
only essential for exploring spintronics devices \cite{Alomar2015} but also
important for developments of spin caloritronics \cite{Bauer2012}. For Rashba
2DES, based on the relaxation time approximation (RTA)
\cite{Tsutsui2012,Islam2012,Wu2014b} in the semiclassical Boltzmann equation
(SBE) approach, it has been suggested recently that the dimensional reduction
of the electronic structure from 2D to 1D can result in enhancements of the
diffusive thermopower and thermoelectric figure of merit \cite{Wu2014b}
$ZT=\left(  \alpha/\sigma\right)  ^{2}\sigma T/\kappa$. Here $\sigma$,
$\alpha$, $\kappa$ and $T$\ denote the electrical conductivity, Peltier
coefficient, thermal conductivity, absolute temperature, respectively. In the
RTA, the enhancement of thermopower was attributed solely to the 1D-like DOS
below the BCP \cite{Wu2014b}.

However, so far, no fully satisfactory theoretical study on thermoelectric
transport exists for the case that $E_{F}$ lies in the vicinity of or below
the band crossing point. This is because that the RTA is inappropriate for
Rashba 2DES at low temperatures when the electron-impurity scattering
dominates, in the case of strong SOC \cite{Ref-2}. For Fermi energies above
the BCP, the RTA can not handle the difference in the relative importance
between the interband and intraband elastic scatterings. This difference is
significant when $E_{F}$ lies near the BCP, so the RTA is unsuitable for this
case. While for Fermi energies below the BCP, the nontrivial FS topology
induces not only the 1D-like DOS, but also nonconventional intraband
scattering \cite{Ref-2}(inter-branch and intra-branch scatterings, the two
branches are denoted in Fig. 1) which is also beyond the scope of RTA. It has
been shown that \cite{Ref-2}, when $E_{F}$ is below the BCP, the
nonequilibrium spin polarization calculated by an exact transport time
solution of the SBE is quite different from the result obtained by the
constant RTA \cite{Dyrdal2013}. This motivates us to employ the exact solution
\cite{Ref-2} of the SBE to systematically study the effects of strong Rashba
SOC on the spin-related thermoelectric transport in 2DES at low temperatures.

In this paper the exact solution of the SBE is employed to calculate
thermoelectric transport coefficients and the figure of merit in Rashba 2DES.
This solution is suitable when electron-impurity scatterings dominate. We show
that the electrical conductivity takes a Drude form when $E_{F}$ is above the
BCP, but a non-Drude form which is a quadratic function of $E_{F}$ when
$E_{F}$ is below the BCP. We found that, the $E_{F}$-dependence of the Peltier
coefficient is not monotonic and the Mott relation \cite{Ziman1972} breaks
down in the vicinity of the BCP. It is shown that the thermopower and
thermoelectric figure of merit have strong enhancements when $E_{F}$ is
tuned below the BCP. This enhancement in the thermoelectric performance is a
combined result of the 1D-like DOS and the unconventional intraband scattering
induced by the nontrivial FS topology in the band valley.

The paper is organized as follows. We present the characteristic properties of
Rashba 2DES in Sec. II. The thermoelectric transport coefficients are
calculated in Sec. III. In Sec. IV, the thermoelectric figure of merit is
presented. The conclusions of the present paper are given in Sec. V.

\section{Characteristic properties of Rashba spintronic systems}

\subsection{The DOS and the topological features of FS}

We consider a Rashba 2DES with spin-independent disorder%
\begin{equation}
H=\frac{\mathbf{p}^{2}}{2m}+\frac{\beta}{\hbar}\mathbf{\hat{\sigma}}\cdot\left(
\mathbf{p}\times\mathbf{\hat{z}}\right)  +V\left(  \mathbf{r}\right)  ,
\end{equation}
where $\mathbf{p=\hbar k}$ is the momentum of the electron, $m$ is the
effective mass, $\mathbf{\hat{\sigma}}=\left(\sigma_{x},\sigma_{y},\sigma
_{z}\right)$ are the Pauli matrices, $\beta$ the Rashba coefficient. The
disorder potential $V\left(\mathbf{r}\right)=\sum_{i}V_{0}\delta\left(
\mathbf{r}-\mathbf{R}_{i}\right)$ is produced by randomly distributed identical
$\delta$-scatters at $\mathbf{R}_{i}$ and is modeled by the standard Gaussian
disorder average $\left\langle \left\vert V_{\mathbf{k}^{\prime}\mathbf{k}%
}\right\vert ^{2}\right\rangle _{dis}=n_{im}V_{0}^{2}$ where $n_{im}$ is the
impurity concentration, $V_{0}$ is the strength of the disorder potential,
$V_{\mathbf{k}^{\prime}\mathbf{k}}$ is the orbital disorder matrix element and
$\left\langle ..\right\rangle _{dis}$ the disorder average.\ The inner
eigenstates and eigenenergies of the clean system read
$|u_{\lambda\mathbf{k}}\rangle=\frac{1}{\sqrt{2}}\left(1,-i\lambda
\exp\left(i\phi\right) \right)^{T}$ and $E_{\lambda k}=\frac{\hbar
^{2}k^{2}}{2m}+\lambda\beta k$, respectively. Here $\lambda=\pm$ and $\phi$ is the polar
angle of $\mathbf{k}$.

The DOS\quad at energy $E\geq0$ is given by $N_{>}\left(  E\right)
=\sum_{\lambda}N_{\lambda}\left(E\right)=2N_{0}$ with $N_{\lambda}\left(
E\right)=N_{0}\frac{k_{\lambda}\left(E\right)}{k_{\lambda}\left(
E\right)+\lambda k_{R}}$ the DOS in the $\lambda$ band. Here $k_{R}%
=m\frac{\beta}{\hbar^{2}}$, $N_{0}=\frac{m}{2\pi\hbar^{2}}$, $k_{\lambda
}\left(E\right)=-\lambda k_{R}+\frac{1}{\beta}\sqrt{E_{R}^{2}+2E_{R}E}$ is
the wave number corresponding to a given energy $E\geq0$ in the $\lambda$
band, $E_{R}=m\left(\frac{\beta}{\hbar}\right)^{2}$ is the "Rashba energy".

\begin{figure}[ptbh]
\includegraphics[width=0.45\textwidth]{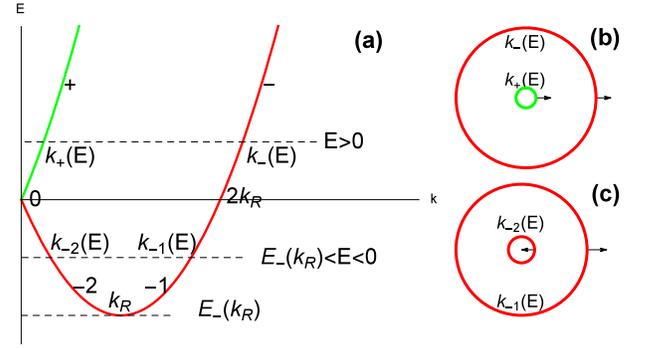} \caption{Band structure
of the Rashba 2DES. (a) Dispersion curve. The energy of the bottom of the
dispersion curve is $E_{-}\left(k_{R}\right)=-\frac{1}{2}E_{R}$ with
$k_{R}=m\frac{\beta}{\hbar^{2}}$. At a given energy $E\geq0$, the wave number
in $\pm$ band is defined as $k_{\pm}\left(E\right)$. For $-\frac{1}%
{2}E_{R}<E\leq0$, there are two monotonic regions on $E-k$ curve: the one from
$k=0$ to $k_{R}$ is marked by the branch $-2$, whereas the other from
$k=k_{R}$ to $2k_{R}$ marked by branch $-1$. The wave number $k_{-\nu}\left(
E\right)$ represents the wave number in the $-\nu$ branch at given $E$,
where $\nu=1,2$. (b) Constant-energy circles for $E>0$. (c) Constant-energy
circles for $-\frac{1}{2}E_{R}<E<0$. The arrows in (b) and (c) represent the
directions of the group velocity.}%
\label{fig1}%
\end{figure}

Below the BCP there is a valley structure in the lower Rashba band, with the
bottom located at $k_{R}$ and the minimal energy $E_{-}\left(k_{R}\right)
=-\frac{1}{2}E_{R}$. At a given energy $E$ in the band valley above the
bottom, there are two wave numbers $k_{-\nu}\left(E\right)=k_{R}+\left(
-1\right)^{\nu-1}\frac{1}{\beta}\sqrt{E_{R}^{2}+2E_{R}E}$ where $\nu=1,2$
denote the two monotonic branches as shown in Fig. 1. It is worth noticing
that the group velocity in the $-2$ branch is anti-parallel to the momentum.
The DOS in this regime has a 1D behavior $N_{<}\left(E\right)=\sum_{\nu
=1}^{2}N_{-\nu}\left(E\right)=2N_{0}\frac{E_{R}}{\sqrt{E_{R}^{2}+2E_{R}E}%
}$ with $N_{-\nu}\left(E\right)=N_{0}\frac{k_{-\nu}\left(E\right)
}{\left\vert k_{-\nu}\left(E\right)-k_{R}\right\vert }$.

\subsection{The response to the external electric field and temperature
gradient}

In the linear response regime, within the semiclassical Boltzmann framework
the electric current density $\mathbf{j}^{e}$\ and heat current density
$\mathbf{j}^{h}$\ are given by%
\begin{equation}
\mathbf{j}^{e}=e\sum_{l}f_{l}\mathbf{v}_{l}\equiv\sigma\mathbf{E}^{\ast
}\mathbf{+}\alpha\left(-\mathbf{\nabla}T\right) \label{transport1}%
\end{equation}
and%
\begin{equation}
\mathbf{j}^{h}=\sum_{l}f_{l}\mathbf{v}_{l}\left(  E_{l}-\mu\right)  \equiv
T\alpha\mathbf{E}^{\ast}\mathbf{+}\kappa\left(  -\mathbf{\nabla}T\right)  .
\label{transport2}%
\end{equation}
Here $l$ is the eigenstate index, $\mathbf{v}_{l}$ is the group velocity of
state $l$, $f_{l}$ denotes the semiclassical distribution function (DF) for
the electron wave packets, $\mathbf{E}^{\ast}=$ $\mathbf{E}-\frac{1}%
{e}\mathbf{\nabla}\mu$ the gradient of the electrochemical potential,
$\mathbf{\nabla}T$ the temperature gradient. $f_{l}=f^{0}\left(E_{l}\right)
+g_{l}$ where $f^{0}$ is the equilibrium Fermi-Dirac DF and $g_{l}$ the
out-of-equilibrium deviation linear in the generalized driven force. The
system is time-reversal invariant, thus the Hall transport is absent and
$\sigma$, $\alpha$, $\kappa$ are all numbers.\ To calculate these
thermoelectric transport coefficients in low temperature cases where the
static impurity scattering dominates, we employ the SBE \cite{Ziman1972}. In
the presence of weak uniform electric field and gradients of chemical
potential and temperature, the linearized SBE suitable for the present system
in nonequilibrium stationary state takes the form \cite{Ziman1972,Vyborny2009}%
:%
\begin{equation}
\mathbf{F}_{l}\cdot\mathbf{v}_{l}\frac{\partial f^{0}}{\partial E_{l}}%
=-\sum_{l^{\prime}}\omega_{l^{\prime},l}\left[  g_{l}-g_{l^{\prime}}\right]  .
\label{SBE}%
\end{equation}
Here $\mathbf{F}_{l}=e\mathbf{E}^{\ast}+\frac{E_{l}-\mu}{T}\left(
-\mathbf{\nabla}T\right)$ is the generalized force acting on state $l$.
$\omega_{l^{\prime},l}$ is the elastic scattering rate from eigenstate $l$ to
$l^{\prime}$, which can be obtained by the golden rule in quantum mechanical
scattering theory. For the present system, the Born approximation in the
lowest order is sufficient \cite{Sinitsyn2008,Vyborny2009}, i.e.,
$\omega_{l^{\prime},l}=\frac{2\pi}{\hbar}\left\langle \left\vert
V_{\mathbf{k}^{\prime}\mathbf{k}}\right\vert ^{2}\right\rangle _{dis}%
\left\vert \langle u_{l^{\prime}}|u_{l}\rangle\right\vert ^{2}\delta\left(
E_{l}-E_{l\prime}\right)$.

The SBE in isotropic 2D Rashba system can be solved conveniently using energy
$E$ and polar angle $\phi$\ and band index $\lambda$ as variables, i.e.,
$l=\left(E,\lambda,\phi\right)$. While the valley region of the lower band
is worth noticing due to the non-monotonic band dispersion. In the valley
region, the branch index $-\nu$ introduced above is needed to denote the
eigenstate, i.e., $l=\left(E,-\nu,\phi\right)$. Then the exact solution of
Eq. (\ref{SBE}) can be obtained, which we refer to our previous work
\cite{Ref-2}:
\begin{equation}
g_{\lambda}\left(  E\right)  =\left(  -\partial_{E}f^{0}\right)
\mathbf{F}_{E}\cdot\frac{\hbar\mathbf{k}_{\lambda}\left(  E\right)  }{m}%
\tau\label{DF1}%
\end{equation}
when $E>0$, and%
\begin{equation}
g_{-\nu}\left(  E\right)  =\left(  -\partial_{E}f^{0}\right)  \mathbf{F}%
_{E}\cdot\frac{\hbar\mathbf{k}_{-\nu}\left(  E\right)  }{m}\tau\left(
-1\right)  ^{\nu-1}\frac{E_{R}^{2}+2E_{R}E}{E_{R}^{2}} \label{DF2}%
\end{equation}
when $E_{-}\left(k_{R}\right)<E<0$. Here $\mathbf{F}_{E}$ represents the
generalized force acting on electrons with energy $E$, $\tau=\left(
\frac{2\pi n_{im}V_{0}^{2}N_{0}}{\hbar}\right)^{-1}$ is the ordinary
momentum relaxation time. At $E=0$, $g_{+}\left(E\rightarrow0^{+}\right)
=g_{-2}\left(E\rightarrow0^{-}\right)=0$, $g_{-1}\left(E\rightarrow
0^{-}\right)=g_{-}\left(E\rightarrow0^{+}\right)$. The DOS at $\left(
E=0,k=0\right)$ vanishes, so this point does not contribute to transport
quantities in Eq. (\ref{transport1}) and (\ref{transport2}), and only the
outer constant-energy circle $\left(E=0,k=2k_{R}\right)$ contributes at
$E=0$.

We note that Eq. (\ref{DF1}) takes into account the intraband ($\lambda
\rightarrow\lambda$) and interband ($\lambda\rightarrow-\lambda$) elastic
scatterings. While Eq. (\ref{DF2}) takes into account the unconventional
intraband scattering in the band valley, i.e., the intra-branch ($-\nu
\rightarrow-\nu$) and inter-branch ($-\nu\rightarrow-(3-\nu)$) elastic scatterings.

\subsection{The chemical potential at low temperatures}

The relation between $E_{F}$ and $\mu$ is needed for investigating the $E_{F}%
$-dependence of Peltier coefficient and thermal conductivity. It can be
obtained by considering the electron density as follows.\textbf{ }At finite
temperatures, the electron density can be calculated by $n_{e}\equiv n_{e}%
^{>}+n_{e}^{<}$, where we define $n_{e}^{>}$ and $n_{e}^{<}$ as%
\begin{align}
n_{e}^{>}  &  =\int_{0}^{\infty}dEf^{0}\left(  E\right)  N_{>}\left(
E\right)  ,\\
n_{e}^{<}  &  =\int_{E_{-}\left(  k_{R}\right)  }^{0}dEf^{0}\left(  E\right)
N_{<}\left(  E\right)  .\nonumber
\end{align}

By defining two functions:%
\begin{equation}
Q_{>}\left(  E\right)  =\int_{0}^{E}dE^{\prime}N_{>}\left(  E^{\prime}\right)
=2N_{0}E
\end{equation}
for $E\geq0$ and%
\begin{equation}
Q_{<}\left(  E\right)  =\int_{E}^{0}dE^{\prime}N_{<}\left(  E^{\prime}\right)
=2N_{0}E_{R}\left(  1-\sqrt{1+2\frac{E}{E_{R}}}\right)
\end{equation}
for $E\leq0$, $n_{e}^{>}$ and $n_{e}^{<}$ can be written in the following
forms via integration by parts:%
\begin{align}
n_{e}^{>}  &  =\int_{0}^{\infty}dE\left(  -\frac{\partial f^{0}}{\partial
E}\right)  Q_{>}\left(  E\right)  ,\label{n>n<}\\
n_{e}^{<}  &  =Q_{<}\left(  E_{-}\left(  k_{R}\right)  \right)  -\int
_{E_{-}\left(  k_{R}\right)  }^{0}dE\left(  -\frac{\partial f^{0}}{\partial
E}\right)  Q_{<}\left(  E\right)  .\nonumber
\end{align}
Setting%

\begin{equation}
\frac{E-\mu}{k_{B}T}=x,\frac{\mu}{k_{B}T}=-t_{1},\frac{E_{F}}{k_{B}T}=-t_{2},
\end{equation}
in Eq. (\ref{n>n<}) and restricting to not too low chemical potential
$\frac{\mu-E_{-}\left(  k_{R}\right)  }{k_{B}T}\gg1$, the total electron
density is found as%
\begin{align}
&  \frac{n_{e}}{2N_{0}E_{R}}=\int_{t_{1}}^{\infty}dx\left(  -\frac{\partial
f^{0}}{\partial x}\right)  \left[  \left(  x-t_{1}\right)  \frac{k_{B}T}%
{E_{R}}+1\right] \nonumber\\
&  +\int_{-\infty}^{t_{1}}dx\left(  -\frac{\partial f^{0}}{\partial x}\right)
\sqrt{1+2\left(  x-t_{1}\right)  \frac{k_{B}T}{E_{R}}}. \label{chemi1}%
\end{align}
Here the condition $\frac{\mu-E_{-}\left(k_{R}\right)}{k_{B}T}\gg1$ can be
realized at low temperatures due to the giant Rashba SOC, e.g., in BiTeI
surface state \cite{Rusinov2013} $E_{-}\left(k_{R}\right)\simeq-90meV$, if
$\mu=\frac{1}{2}E_{-}\left(k_{R}\right)$ we have $\frac{\mu-E_{-}\left(
k_{R}\right)}{k_{B}}\simeq450K$. At zero-temperature $\mu=E_{F}$, when
$E_{F}\geq0$ it has been obtained that \cite{Novokshonov2006} $n_{e}\left(
E_{F}\geq0\right)=2N_{0}\left(E_{F}+E_{R}\right)$. While when $E_{F}%
\leq0$ we obtain%
\begin{equation}
n_{e}\left(  E_{F}\leq0\right)  =\int_{k_{-2}\left(  E_{F}\right)  }%
^{k_{-1}\left(  E_{F}\right)  }\frac{kdk}{2\pi}=2N_{0}\sqrt{E_{R}^{2}%
+2E_{R}E_{F}}. \label{density<}%
\end{equation}
In this case the Fermi level intersects only the lower band and only the
annulus lying between the two Fermi circles of radii $k_{-1,-2}\left(
E_{F}\right)$\ is filled. This nontrivial topology of Fermi surfaces in the
band valley\ has been highlighted in previous researches
\cite{Tsutsui2012,Cappelluti2007,Lv2013}. Substituting the electron density
into Eq. (\ref{chemi1}), we obtain the Fermi energy dependence of chemical
potential at a given temperature, as presented in Fig. 2. \begin{figure}[ptbh]
\includegraphics[width=0.4\textwidth]{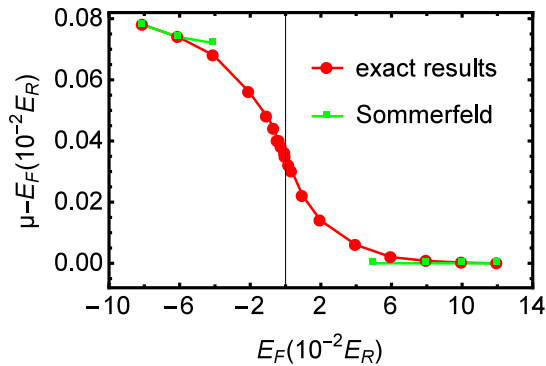} \caption{The
difference between the chemical potential and the Fermi energy. The
temperature is fixed to $0.02E_{R}/k_{B}$. The numerical results describe
exactly the behavior near $E_{F}=0$, which is beyond the scope of the
Sommerfeld expansion.}%
\label{fig2}%
\end{figure}

When $E_{F}\geq0.1E_{R}$, $\mu-E_{F}=0$, same as the analytic result based on
the Sommerfeld expansion \cite{Ziman1972} in the case of\ $E_{F}\gg k_{B}T$.
When $E_{F}\leq-0.08E_{R}$, numerical results can be well fitted by the
formula $\mu-E_{F}=\frac{\pi^{2}}{6}\frac{\left(k_{B}T\right)^{2}}%
{E_{R}+2E_{F}}$ obtained by the Sommerfeld expansion in the case of $-E_{F}\gg
k_{B}T$ and $E_{F}+\frac{1}{2}E_{R}\gg k_{B}T$ (this latter condition ensures
that the band valley structure and Fermi surface are not smeared out by
thermal broadening and can be satisfied at low temperatures with giant Rashba
SOC). It is well-known that at low temperatures in three-dimensional (3D)
parabolic 2DES $\mu<E_{F}$, in 2D $\mu=E_{F}$ and in 1D $\mu>E_{F}$, due to
different energy dependencies of the DOS \cite{Ziman1972}. Here the above two
limiting cases correspond to 2D and 1D cases, due to the 2D- and 1D-like DOS
for energies above and below the BCP, respectively.

When $E_{F}$ is in the intermediate region $-0.08E_{R}\leq E_{F}\leq0.1E_{R}$,
the Sommerfeld expansion is not valid and numerical results clearly show
continuous transition between above two limiting cases.\ This transition from
2D to 1D is resulted by the Fermi surfaces topology change in Rashba model.

\section{Thermoelectric response coefficients}

\subsection{Drude-like and non-Drude forms of electrical conductivities}

In what follows we assume that the generalized force is applied in $x$
direction. When both bands are partially occupied, the zero-temperature
electrical conductivity is calculated by substituting the group velocity
$\mathbf{v}\left(E,\lambda,\phi\right)=\frac{N_{0}}{N_{\lambda}\left(
E\right)}\frac{\hbar\mathbf{k}_{\lambda}\left(E\right)}{m}$ and the DF
Eq. (\ref{DF1})\ into Eq. (\ref{transport1}). The result is
\begin{equation}
\sigma\left(  E_{F}\geq0\right)  =\frac{e^{2}}{2\pi^{2}\hbar}\frac{2\pi\left(
E_{F}+E_{R}\right)  \tau}{\hbar}, \label{Drude}%
\end{equation}
where and below we use the notation $\sigma\left(E_{F}\right)$ to
represent the zero-temperature electrical conductivity $\sigma\left(
T=0,E_{F}\right)$ for brevity.

This result still has the usual form of Drude formula $\sigma=n_{e}e^{2}%
\tau/m$. It has been obtained in some earlier works based on the Green's
function calculation in the ladder approximation \cite{Novokshonov2006} or the
same exact solution to the SBE \cite{Trushin2007} as our Eq. (\ref{DF1}) when
$E_{F}\geq0$. In this case it looks as if the electric current were generated
by charge carriers of one type with density $n_{e}$ and mobility $e\tau/m$.

For Fermi energies below the BCP, substituting $g_{-\nu}$ and the group
velocity $\mathbf{v}\left(E,-\nu,\phi\right)=\left(-1\right)^{\nu
-1}\frac{N_{0}}{N_{-\nu}\left(E\right)}\frac{\hbar\mathbf{k}_{-\nu}\left(
E\right)}{m}$ into Eq. (\ref{transport1}), we obtain the zero-temperature
electrical conductivity
\begin{equation}
\sigma\left(  E_{F}\leq0\right)  =\frac{e^{2}}{2\pi^{2}\hbar}\frac{2\pi\left(
E_{F}+E_{R}\right)  \tau}{\hbar}\frac{E_{R}^{2}+2E_{R}E_{F}}{E_{R}^{2}}.
\label{non-Drude}%
\end{equation}
It has a quadratic dependence on the Fermi energy and does not take the form
of Drude conductivity, different from Eq. (\ref{Drude}). Since the Fermi
surfaces topology in the band valley differs from that above the BCP, the
behaviors of electrical conductivities are different between the two regions.

\subsection{The Peltier coefficient}

The Peltier coefficient and the electrical conductivity are connected by
(details in the Appendix)%
\begin{equation}
\alpha=\frac{1}{e}\int_{E_{-}\left(  k_{R}\right)  }^{\infty}dE\left(
-\frac{\partial f^{0}}{\partial E}\right)  \frac{E-\mu}{T}\sigma\left(
E\right)  , \label{ther-elec}%
\end{equation}
where $\sigma\left(E\right)$ represents the zero-temperature electrical
conductivity with Fermi energy $E$. Usually, this energy integration is worked
out by performing the Sommerfeld expansion \cite{Ziman1972}. When the
Sommerfeld expansion is valid, the Mott relation holds. The former demands
that $\sigma\left(E\right)$ is continuously differentiable at the chemical
potential. However, according to Eqs. (\ref{Drude}) and (\ref{non-Drude}),
$\sigma\left(E\right)$ takes different forms between the two sides of the
BCP. Therefore $\frac{\partial\sigma\left(E\right)}{\partial E}|_{\mu}$ is
not continuous and the Sommerfeld expansion is not valid at $\mu=0$, thus the
Mott relation fails for chemical potentials near the BCP.

\begin{figure*}[ptbh]
\includegraphics[width=0.84\textwidth]{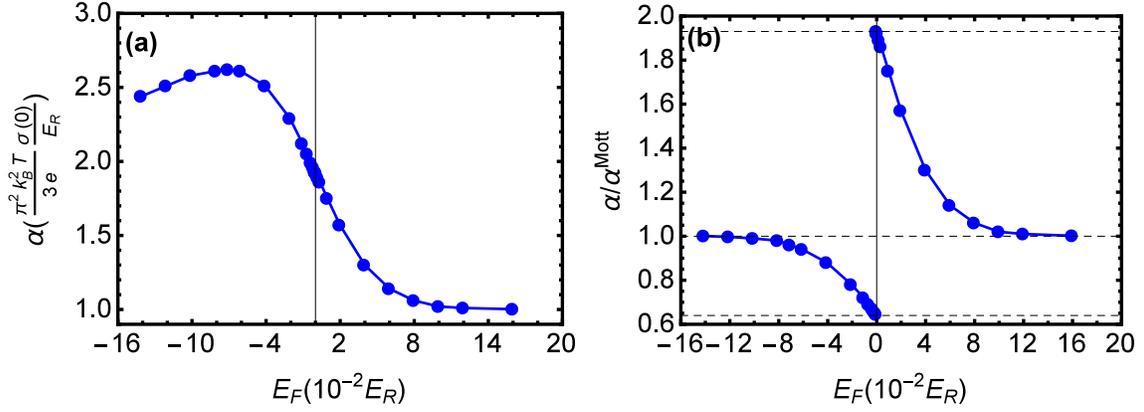} \caption{(a) The Fermi
energy dependence of the Peltier Coefficient. (b) The comparison between the
numerical results and the Mott relation. The deviation from the Mott relation
is significant when the Fermi energy lies near the BCP $E_{F}=0$. In (a) and
(b), the temperature is fixed to $0.02E_{R}/k_{B}$. The relation between the
chemical potential and Fermi energy shown in figure 2 has been taken into
account}%
\label{fig3}%
\end{figure*}

Substituting Eqs. (\ref{Drude}) and (\ref{non-Drude}) into Eq.
(\ref{ther-elec}),\ we obtain the Peltier coefficient as%
\begin{align}
\alpha &  =\frac{\pi^{2}k_{B}^{2}T}{3e}\frac{\sigma\left(  0\right)  }{E_{R}%
}3\left\{  1-2\frac{b\left(  t_{1}\right)  -t_{1}a\left(  t_{1}\right)  }%
{\pi^{2}}\right. \nonumber\\
&  \left.  -\frac{2}{\pi^{2}}\left[  c\left(  t_{1}\right)  +2t_{1}\left(
\frac{\pi^{2}}{3}-b\left(  t_{1}\right)  \right)  +t_{1}^{2}a\left(
t_{1}\right)  \right]  \frac{k_{B}T}{E_{R}}\right\}  , \label{main}%
\end{align}
where we define%
\begin{align}
a\left(  t_{1}\right)   &  =\int_{t_{1}}^{\infty}dx\left(  -\frac{\partial
f^{0}}{\partial x}\right)  x,\nonumber\\
b\left(  t_{1}\right)   &  =\int_{t_{1}}^{\infty}dx\left(  -\frac{\partial
f^{0}}{\partial x}\right)  x^{2},\\
c\left(  t_{1}\right)   &  =\int_{t_{1}}^{\infty}dx\left(  -\frac{\partial
f^{0}}{\partial x}\right)  x^{3}.\nonumber
\end{align}
Eq. (\ref{main}) can describe the behavior of $\alpha$ for Fermi energies near
the BCP. Because the Sommerfeld expansion is unsuitable in the band crossing
region, we have to perform numerical calculations.

For temperature at $0.02E_{R}/k_{B}$, the $E_{F}$-dependence of Peltier
coefficient is shown in Fig. 3(a). We find that when $E_{F}/E_{R}\geq0.1$,
$\alpha$ is almost constant with respect to $E_{F}$, which is consistent with
the Mott relation (Eq. (\ref{Mott>})); when $E_{F}/E_{R}\leq-0.08$, $\alpha$
has a nearly linear dependence on $E_{F}$, which is also consistent with the
Mott relation (Eq. (\ref{Mott<})).\ Between above two regions, a non-monotonic
Fermi energy dependence of $\alpha$ in the band crossing region is found.

We use the notation $\alpha^{Mott}$ to denote the Peltier coefficient obtained
by the Mott relation $\alpha^{Mott}=\frac{\pi^{2}k_{B}^{2}T}{3e}\frac
{\partial\sigma\left(E\right)}{\partial E}|_{E=E_{F}}$:
\begin{equation}
\alpha^{Mott}=\frac{\pi^{2}k_{B}^{2}T}{3e}\frac{\sigma\left(  0\right)
}{E_{R}},E_{F}\geq0, \label{Mott>}%
\end{equation}
and%
\begin{equation}
\alpha^{Mott}=\frac{\pi^{2}k_{B}^{2}T}{3e}\sigma\left(  0\right)  \frac
{3}{E_{R}}\left[  1+\frac{4}{3}\frac{E_{F}}{E_{R}}\right]  ,E_{F}\leq0.
\label{Mott<}%
\end{equation}
Combining Eq. (\ref{main}) with Eqs. (\ref{Mott>}) and (\ref{Mott<}), we plot
$\alpha/\alpha^{Mott}$ in Fig. 3(b). When $-0.08E_{R}\leq E_{F}\leq0.1E_{R}$,
our results show deviations from the Mott relation.

The Mott relation fails near the BCP due to the fact that the electrical
conductivity takes different forms on the two sides of the BCP. Therefore the
deviation from Mott relation can be regarded as a consequence of the
topological change of FS varying across the BCP.

\subsection{The thermal conductivity}

The thermal current response to the temperature gradient can be obtained by
(details in the Appendix)%
\begin{equation}
\kappa=\left(  \frac{k_{B}}{e}\right)  ^{2}T\int_{E_{-}\left(  k_{R}\right)
}^{\infty}dE\left(  -\frac{\partial f^{0}}{\partial E}\right)  \left(
\frac{E-\mu}{k_{B}T}\right)  ^{2}\sigma\left(  E\right)  . \label{ther cond}%
\end{equation}
Substituting Eqs. (\ref{Drude}) and (\ref{non-Drude}) into Eq.
(\ref{ther cond}), we obtain%
\begin{align}
&  \frac{\kappa}{L_{0}\sigma\left(  0\right)  T}=1-\frac{2c\left(
t_{1}\right)  +t_{1}\left(  \pi^{2}-2b\left(  t_{1}\right)  \right)  }{\pi
^{2}/3}\frac{k_{B}T}{E_{R}}\nonumber\\
&  +\frac{\frac{7\pi^{4}}{15}-d\left(  t_{1}\right)  +2t_{1}c\left(
t_{1}\right)  +t_{1}^{2}\left(  \frac{\pi^{2}}{3}-b\left(  t_{1}\right)
\right)  }{\pi^{2}/6}\left(  \frac{k_{B}T}{E_{R}}\right)  ^{2}, \label{main1}%
\end{align}
where $L_{0}=\frac{1}{3}\left(  \frac{\pi k_{B}}{e}\right)  ^{2}$ is the
free-electron Lorentz number, and $d\left(  t_{1}\right)  =\int_{t_{1}%
}^{\infty}dx\left(  -\frac{\partial f^{0}}{\partial x}\right)  x^{4}$. This is
our main result for $\kappa$.

\begin{figure*}[ptbh]
\includegraphics[width=0.84\textwidth]{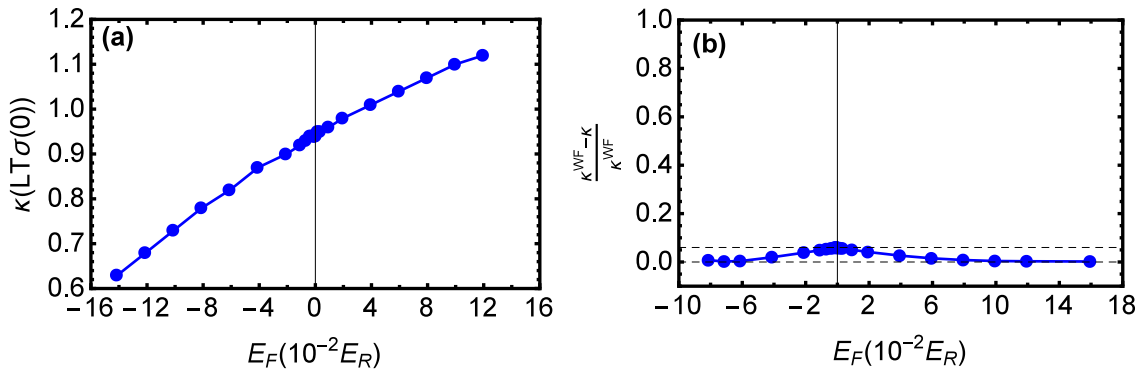}
\caption{(a) The Fermi energy dependence of the thermal conductivity. (b) The
comparison between the numerical results and the Wiedemann-Franz law. The
deviation is very slight, about $<6\%$ near the BCP. In (a) and (b), the
temperature is fixed to $0.02E_{R}/k_{B}$.}%
\label{fig4}%
\end{figure*}

The $E_{F}$-dependence of $\kappa$ is given in Fig. 4(a). It shows monotonic
dependence on the Fermi energy of $\kappa$.

We compare our numerical results with that given by the Wiedemann-Franz law in
Fig. 4(b). Here $\kappa^{WF}$ denotes the thermal conductivity based on the
Wiedemann-Franz law $\kappa^{WF}=LT\sigma\left(\mu\right)\simeq
LT\sigma\left(E_{F}\right)$. The deviation from Wiedemann-Franz law is
very slight: $\left\vert \frac{\kappa-\kappa^{WF}}{\kappa^{WF}}\right\vert
<6\%$ in the whole regime we investigate here. Hence the Wiedemann-Franz law
holds quite well. The difference between the validity of the Mott relation and
the Wiedemann-Fran law in the band crossing region can be understood as
follows. The Mott relation connects the Peltier coefficient with the energy
derivative of the electrical conductivity, while the Wiedemann-Franz law
connects the thermal conductivity with the electrical conductivity itself. At
the BCP, the electrical conductivity is continuous, but its energy derivative
is not. The Peltier coefficient based on the Mott relation is not continuous
across BCP. Therefore, in the vicinity of the BCP the Mott relation breaks
down, while the Wiedemann-Franz law is satisfied quite well.

\subsection{Comparison with RTA results}

Eqs. (\ref{Drude}) and (\ref{non-Drude}) are different from the electrical
conductivities obtained by employing the constant RTA \cite{Tsutsui2012}:%
\begin{align}
\sigma^{RTA}\left(  E_{F}\geq0\right)   &  =\frac{e^{2}}{2\pi^{2}\hbar}%
\frac{2\pi\left(  E_{F}+\frac{1}{2}E_{R}\right)  \tau}{\hbar},\nonumber\\
\sigma^{RTA}\left(  E_{F}\leq0\right)   &  =\frac{e^{2}\tau}{2\pi\hbar^{2}%
}\sqrt{E_{R}^{2}+2E_{R}E_{F}}. \label{RTA}%
\end{align}
The RTA result neither takes the Drude form for $E_{F}\geq0$ nor has a
polynomial dependence on $E_{F}/E_{R}$ for $E_{F}\leq0$.\ When $E_{F}\geq0$,
$\sigma^{RTA}$ tends to Eq. (\ref{Drude}) for weak SOC $E_{F}\gg E_{R}$. For
Fermi energies near the BCP, the electrical conductivity obtained by the
constant RTA significantly differs from Eqs. (\ref{Drude}) and
(\ref{non-Drude}): $\sigma\left(E_{F}=0\right)=n_{e}\left(E_{F}%
=0\right)e^{2}\tau/m=2\sigma^{RTA}\left(E_{F}=0\right)$. For $E_{F}$
below the BCP, as long as the band valley structure can survive the thermal
smearing and disorder broadening: $E_{R}\gg k_{B}T$, $\hbar/\tau$, the
difference between Eq. (\ref{non-Drude}) and $\sigma^{RTA}$ can not be ignored.

These differences between our results and the RTA results can be understood as
follows. When $E_{F}$ lies high above the BCP $E_{F}\gg E_{R}$, $N_{+}\left(
E_{F}\right)\simeq N_{-}\left(E_{F}\right)$, the intraband and interband
scattering events are of equal importance for both the inner (+) and outer (-)
Fermi circles. Only in this case the RTA works well. When $E_{F}$ lies near
the BCP, $N_{+}\left(E_{F}\right)$ tends to zero. Thus, for Fermi
electrons on the inner Fermi circle the interband scattering events dominate
over the intraband scattering \cite{Ye2015}; while for Fermi electrons on the
outer Fermi circle the intraband scattering events dominate over the interband
scattering. The difference in the relative importance between the intraband
and interband scattering events as well as its change with varying $E_{F}$ can
not be described by the RTA. When $E_{F}$ lies below the BCP, the non-Drude
form of $\sigma^{RTA}\left(E_{F}<0\right)  $ is caused only by the 1D-like
DOS below the BCP, while that of Eq. (\ref{non-Drude}) based on the exact
solution of SBE relies on not only the 1D-like DOS but also the unconventional
intraband scattering (inter-branch and intra-branch scatterings) induced by
the nontrivial FS topology.

Now we examine the thermopower in the "Mott relation regimes": $S=\alpha
^{Mott}/\sigma$. When $E_{F}\gg k_{B}T$%
\begin{equation}
S=\frac{\pi^{2}k_{B}^{2}T}{3e}\frac{1}{E_{F}+E_{R}}=\frac{\pi^{2}k_{B}^{2}%
T}{3e}\frac{2N_{0}}{n_{e}} \label{TEP>}%
\end{equation}
and when $-E_{F}\gg k_{B}T$%
\begin{equation}
S=\frac{\pi^{2}k_{B}^{2}T}{3e}\frac{3}{E_{R}}\frac{\ 1+\frac{4}{3}\frac{E_{F}%
}{E_{R}}\ }{\left(  1+\frac{E_{F}}{E_{R}}\right)  \left(  1+2\frac{E_{F}%
}{E_{R}}\right)  }. \label{TEP<}%
\end{equation}
The thermopower is enhanced in the band valley, similar to the thermoelectric
figure of merit which will be discussed in the next section (Eqs. (\ref{ZT})).
Eqs. (\ref{TEP>}) and (\ref{TEP<}) are different from those obtained by the
constant RTA \cite{Wu2014b}:%
\begin{align}
S^{RTA}\left(  E_{F}\gg k_{B}T\right)   &  =\frac{\pi^{2}k_{B}^{2}T}{3e}%
\frac{1}{E_{F}+\frac{1}{2}E_{R}},\nonumber\\
S^{RTA}\left(  -E_{F}\gg k_{B}T\right)   &  =\frac{1}{2}\frac{\pi^{2}k_{B}%
^{2}T}{3e}\frac{1}{E_{F}+\frac{1}{2}E_{R}}. \label{TEP-RTA}%
\end{align}
Based on the RTA result Eq. (\ref{TEP-RTA}), it had been concluded that below
the BCP the enhancement of thermopower in a Rashba 2DES compared to a
parabolic 2DES with the same electron density is caused solely by the much
lower $E_{F}$. And this lower Fermi energy is a consequence of the 1D-like DOS
\cite{Wu2014b}. However, Eq. (\ref{TEP<}) obtained using the exact solution of
the SBE shows that the enhancement of thermopower is a combined result of the
1D-like DOS and the unconventional inter-branch and intra-branch scatterings
in the band valley.

In conventional semiconductor asymmetric quantum-wells the difference between
Eq. (\ref{Drude}) and the 1st equation of Eq. (\ref{RTA}) is negligible due to
the weak Rashba SOC $E_{F}\gg E_{R}$, so does the difference between Eq.
(\ref{TEP>}) and the 1st equation of Eq. (\ref{TEP-RTA}). In Rashba
semiconductors BiTeX (X=Cl, Br, I), despite that the reported Fermi levels in
existing experiments are still in 3D bulk conduction band (BCB), the validity
of constant RTA analysis of electrical conductivity has been questioned when
$E_{F}$ lies near the BCP of BCB \cite{Ye2015}. With further studies on
systematic doping in these Rashba semiconductors \cite{Wang2013,Ye2015}, it is
promising that the Fermi level can be tuned into the bulk band gap. In
addition, the BiTeX quantum well \cite{Wu2014b} is another possible candidate
to realize strongly spin-orbit coupled 2DES. Very recently the first-principle
calculation has suggested the formation of 2DES with large Rashba SOC by
strain engineered growth of a Au single layer on the layered large band-gap
semiconductor InSe(0001) substrate \cite{Ming2015}. Future studies of the
strain engineering of heavy-metal film on layered large-gap semiconductor
substrate may also realize 2DES with stronger Rashba SOC. In these systems the
transport properties of the 2DES with strong Rashba SOC can be detected
experimentally and our theoretical results can be tested. For the experimental
measurements of the low-temperature diffusive thermopower, the hot-electron
thermocouple technique could be applied, which is much less sensitive to
phonon-drag effects than conventional methods \cite{Chickering2009}.

\section{The thermoelectric figure of merit and the enhancement below the BCP}

The performance of a thermoelectric material is determined by the figure of
merit $ZT=\left(\alpha/\sigma\right)^{2}\sigma T/\kappa$.

\begin{figure}[ptbh]
\includegraphics[width=0.45\textwidth]{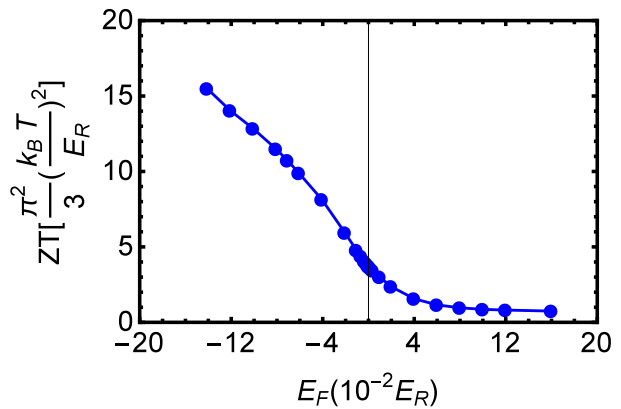} \caption{The Fermi energy
dependence of the figure of merit ZT when the temperature is fixed to
$0.02E_{R}/k_{B}$.}%
\label{fig5}%
\end{figure}For the case that the Fermi energy lies near the BCP, the ZT is
shown in Fig. 5. It shows that when the Fermi energy is tuned across the BCP
from $E_{F}=0.1E_{R}$ down to $E_{F}=-0.08E_{R}$, the figure of merit acquires
a strong enhancement.

When $E_{F}/E_{R}\geq0.1$ or $E_{F}/E_{R}\leq-0.08$, the Mott relation holds
very well, thus $ZT$ can be calculated using $\alpha^{Mott}$, yielding
$ZT=\left(\alpha^{Mott}/\sigma\right)^{2}/L$, i.e.,%
\begin{equation}
ZT=\left\{
\begin{array}
[c]{c}%
\frac{\pi^{2}}{3}\left(  \frac{k_{B}T}{E_{R}}\right)  ^{2}\frac{1}{\left(
1+\frac{E_{F}}{E_{R}}\right)  ^{2}},E_{F}\gtrsim0.1E_{R}\\
\frac{\pi^{2}}{3}\left(  \frac{k_{B}T}{E_{R}}\right)  ^{2}\frac{\left(
3+4\frac{E_{F}}{E_{R}}\right)  ^{2}}{\left(  1+\frac{E_{F}}{E_{R}}\right)
^{2}\left(  1+2\frac{E_{F}}{E_{R}}\right)  ^{2}},E_{F}\lesssim-0.08E_{R}%
\end{array}
\right.  \label{ZT}%
\end{equation}
In both energy intervals, $ZT$ monotonically increases with decreasing Fermi
energy, similar to the case when $-0.08\leq E_{F}/E_{R}\leq0.1$ shown in Fig. 5.

This enhancement in $ZT$ when $E_{F}$ is tuned below the BCP is directly
related to the enhanced Peltier coefficient and the decreased electrical
conductivity when $E_{F}$ varies below the BCP, as shown in Fig. 3(a) and Eqs.
(\ref{Drude}) and (\ref{non-Drude}). Because the Peltier coefficient is
connected with the electrical conductivity , the enhancement in $ZT$ can be
attributed to the nontrivial topology of Fermi surfaces in the band valley regime.

\section{Conclusions}

We have calculated thermoelectric transport coefficients and the figure of
merit in strongly spin-orbit coupled Rashba 2DES with spin independent
disorder using the exact solution of the linearized Boltzmann equation. At low
temperatures the static impurity scattering dominates, it is shown that the
electrical conductivity takes a Drude form when the Fermi energy $E_{F}$ is
above the band crossing point, but a non-Drude form which is a quadratic
function of $E_{F}$ for Fermi energies below the band crossing point. This is
attributed to the different topologies of Fermi surfaces on the two sides of
the band crossing point. For Fermi energies near the band crossing point, the
$E_{F}$-dependence of the Peltier coefficient is not monotonic, and the Mott
relation breaks down. While the thermal conductivity is monotonically
increasing as a function of $E_{F}$ and the Wiedemann-Franz law holds quite
well. The thermopower and figure of merit are strongly enhanced when $E_{F}$ downs
below the band crossing point. This enhancement is caused not only by the
1D-like density of state but also by the unconventional intraband elastic
scattering below the band crossing point.

Our results differ from previous ones obtained by the relaxation time
approximation, especially for Fermi energies in the vicinity of and below the
band crossing point in systems with strong Rashba spin splitting. For Fermi
energies above the band crossing point, our results can handel the difference
in the relative importance between the interband and intraband elastic
scattering events, in contrast to the relaxation time approximation. This
difference is significant when $E_{F}$ lies in the vicinity of the band
crossing point. For Fermi energies below the band crossing point, our results
take into account the unconventional intraband scattering induced by the
nontrivial FS topology, which can not be described by the relaxation time approximation.

Our theoretical results may be tested in strongly spin-orbit coupled 2DES,
e.g., the surface state of polar semiconductors BiTeX (X=Cl, Br, I) and BiTeX
quantum wells, as well as the 2DES formed by the strain engineering of
heavy-metal film on layered large-gap semiconductor substrate \cite{Ming2015}.

\begin{acknowledgments}
The authors are thankful for the support of NSFC (No.11274013 and No.11274018)
and NBRP of China (2012CB921300).
\end{acknowledgments}

\appendix

\section{Derivations of the analytic formulas for the Peltier coefficient and
thermal conductivity}

In order to establish the connection between the electrical conductivity and
Peltier coefficient and thermal conductivity, only the original transport time
form of $g_{l}$ is needed, i.e.,%

\begin{equation}
g_{\lambda}\left(  E\right)  =\left(  -\frac{\partial f^{0}}{\partial
E}\right)  \mathbf{F}_{E}\cdot\mathbf{v}\left(  E,\lambda,\phi\right)
\tau_{\lambda}\left(  E\right)  ,\label{DF>}%
\end{equation}
and%
\begin{equation}
g_{-\nu}\left(  E\right)  =\\
\left(  -\frac{\partial f^{0}}{\partial E}\right)  \mathbf{F}_{E}%
\cdot\mathbf{v}\left(  E,-\nu,\phi\right)  \tau_{-\nu}\left(  E\right)
,\label{DF<}%
\end{equation}
(In this model $\tau_{\lambda}\left(  E\right)  =\tau\frac{N_{\lambda}\left(
E\right)  }{N_{0}}$ and $\tau_{-\nu}\left(  E\right)  =\tau\frac{N_{-\nu
}\left(  E\right)  }{N_{0}}\frac{E_{R}^{2}+2E_{R}E}{E_{R}^{2}}$, which has
been included in and can be read out from the specific form of nonequilibrium
DF in the paper. However, for the purpose in this supplementary material, this
specific form is not needed.) Substituting Eqs. (\ref{DF>}) and (\ref{DF<})
into the two linear response equations, since the generalized force is applied
in $x$ direction, the the electrical conductivity is given by $\sigma
=\sigma^{>}+\sigma^{<}$, where%
\begin{align}
\sigma^{>} &  =e^{2}\int\frac{d\phi}{2\pi}\int_{0}^{\infty}dE\left(
-\frac{\partial f^{0}}{\partial E}\right)  \nonumber\\
&  \times\sum_{\lambda}N_{\lambda}\left(  E\right)  v_{x}^{2}\left(
E,\lambda,\phi\right)  \tau_{\lambda}\left(  E\right)
\end{align}
and%
\begin{align}
\sigma^{<} &  =e^{2}\int\frac{d\phi}{2\pi}\int_{E_{-}\left(  k_{R}\right)
}^{0}dE\left(  -\frac{\partial f^{0}}{\partial E}\right)  \nonumber\\
&  \times\sum_{\nu}N_{-\nu}\left(  E\right)  v_{x}^{2}\left(  E,-\nu
,\phi\right)  \tau_{-\nu}\left(  E\right)  .
\end{align}
And the Peltier coefficient is given by
$\alpha=\alpha^{>}+\alpha^{<}$ with%
\begin{align}
\alpha^{>} &  =e\int\frac{d\phi}{2\pi}\int_{0}^{\infty}dE\left(
-\frac{\partial f^{0}}{\partial E}\right)  \frac{E-\mu}{T}\nonumber\\
&  \times\sum_{\lambda}N_{\lambda}\left(  E\right)  v_{x}^{2}\left(
E,\lambda,\phi\right)  \tau_{\lambda}\left(  E\right)
\end{align}
and%
\begin{align}
\alpha^{<} &  =e\int\frac{d\phi}{2\pi}\int_{E_{-}\left(  k_{R}\right)  }%
^{0}dE\left(  -\frac{\partial f^{0}}{\partial E}\right)  \frac{E-\mu}%
{T}\nonumber\\
&  \times\sum_{\nu}N_{-\nu}\left(  E\right)  v_{x}^{2}\left(  E,-\nu
,\phi\right)  \tau_{-\nu}\left(  E\right).
\end{align}
The thermal conductivity is found as $\kappa=\kappa^{>}+\kappa^{<}$ with%
\begin{align}
\kappa^{>} &  =\int\frac{d\phi}{2\pi}\int_{0}^{\infty}dE\left(  -\frac
{\partial f^{0}}{\partial E}\right)  \frac{\left(  E-\mu\right)  ^{2}}%
{T}\nonumber\\
&  \times\sum_{\lambda}N_{\lambda}\left(  E\right)  v_{x}^{2}\left(
E,\lambda,\phi\right)  \tau_{\lambda}\left(  E\right)
\end{align}
and%
\begin{align}
\kappa^{<} &  =\int\frac{d\phi}{2\pi}\int_{E_{-}\left(  k_{R}\right)  }%
^{0}dE\left(  -\frac{\partial f^{0}}{\partial E}\right)  \frac{\left(
E-\mu\right)  ^{2}}{T}\nonumber\\
&  \times\sum_{\nu}N_{-\nu}\left(  E\right)  v_{x}^{2}\left(  E,-\nu
,\phi\right)  \tau_{-\nu}\left(  E\right)  .
\end{align}
Therefore the Peltier coefficient can be expressed as%
\begin{align}
\alpha^{>} &  =\frac{1}{e}\int_{0}^{\infty}dE\left(  -\frac{\partial f^{0}%
}{\partial E}\right)  \frac{E-\mu}{T}\sigma^{>}\left(  E\right)  \nonumber\\
&  =\frac{1}{e}\int_{0}^{\infty}dE\left(  -\frac{\partial f^{0}}{\partial
E}\right)  \frac{E-\mu}{T}\sigma\left(  E\right)
\end{align}
and%
\begin{align}
\alpha^{<} &  =\frac{1}{e}\int_{E_{-}\left(  k_{R}\right)  }^{0}dE\left(
-\frac{\partial f^{0}}{\partial E}\right)  \frac{E-\mu}{T}\sigma^{<}\left(
E\right)  \nonumber\\
&  =\frac{1}{e}\int_{E_{-}\left(  k_{R}\right)  }^{0}dE\left(  -\frac{\partial
f^{0}}{\partial E}\right)  \frac{E-\mu}{T}\sigma\left(  E\right)  ,
\end{align}
so that%
\begin{equation}
\alpha=\frac{1}{e}\int_{E_{-}\left(  k_{R}\right)  }^{\infty}dE\left(
-\frac{\partial f^{0}}{\partial E}\right)  \frac{E-\mu}{T}\sigma\left(
E\right)  .
\end{equation}
Similarly, the themal conductivity is connected with the zero-temperature
electrical conductivity as
\begin{equation}
\kappa^{>}=\left(  \frac{k_{B}}{e}\right)  ^{2}T\int_{0}^{\infty}dE\left(
-\frac{\partial f^{0}}{\partial E}\right)  \left(  \frac{E-\mu}{k_{B}%
T}\right)  ^{2}\sigma\left(  E\right)
\end{equation}
and%
\begin{equation}
\kappa^{<}=\left(  \frac{k_{B}}{e}\right)  ^{2}T\int_{E_{-}\left(
k_{R}\right)  }^{0}dE\left(  -\frac{\partial f^{0}}{\partial E}\right)
\left(  \frac{E-\mu}{k_{B}T}\right)  ^{2}\sigma\left(  E\right)  ,
\end{equation}
so we have%
\begin{equation}
\kappa=\left(  \frac{k_{B}}{e}\right)  ^{2}T\int_{E_{-}\left(  k_{R}\right)
}^{\infty}dE\left(  -\frac{\partial f^{0}}{\partial E}\right)  \left(
\frac{E-\mu}{k_{B}T}\right)  ^{2}\sigma\left(  E\right)  .
\end{equation}

\end{document}